\documentclass[runningheads]{llncs}
\usepackage{graphicx}
\usepackage{ctable}
\usepackage{framed,multirow}

\begin{document}
\title{Sinc-based convolutional neural networks for EEG-BCI-based motor imagery classification\thanks{This work was supported by MIUR (Minister for Education, University and Research, Law 232/216, Department of Excellence).}}
%
%
\author{Alessandro Bria\inst{1}\orcidID{0000-0002-2895-6544} \and
Claudio Marrocco\inst{1}\orcidID{0000-0003-0840-7350} \and
Francesco Tortorella\inst{2}\orcidID{0000-0002-5033-9323}}
\authorrunning{A. Bria et al.}
\titlerunning{Sinc-based convolutional neural networks for motor imagery classification}
%
\institute{University of Cassino and Southern Latium, Cassino, FR 03043 Italy \and
University of Salerno, Fisciano, SA 84084 Italy}
%
\maketitle              
\begin{abstract}
Brain-Computer Interfaces (BCI) based on motor imagery translate mental motor images recognized from the electroencephalogram (EEG) to control commands. EEG patterns of different imagination tasks, e.g. hand and foot movements, are effectively classified with machine learning techniques using band power features. Recently, also Convolutional Neural Networks (CNNs) that learn both effective features and classifiers simultaneously from raw EEG data have been applied. However, CNNs have two major drawbacks: (i) they have a very large number of parameters, which thus requires a very large number of training examples; and (ii) they are not designed to explicitly learn features in the frequency domain. To overcome these limitations, in this work we introduce Sinc-EEGNet, a lightweight CNN architecture that combines learnable band-pass and depthwise convolutional filters. Experimental results obtained on the publicly available BCI Competition IV Dataset 2a show that our approach outperforms reference methods in terms of classification accuracy.


\end{abstract}
\section{Introduction}
A Brain-Computer Interface (BCI) translates brain signals into messages or commands for an interactive task. This enables a wide range of applications from clinic to industry for both patients and healthy users, such as rehabilitation devices for stroke patients \cite{pichiorri15}, controllable wheelchairs and prostheses \cite{zhang15}, new gaming input devices \cite{coyle2013}, to name a few. Among different brain activity monitoring modalities, noninvasive approaches based on electroencephalography (EEG) use multiple electrodes placed on the skull surface to record the activity of cerebral cortical neurons \cite{britton16} and are widely used in many BCI studies thanks to their ease of implementation, reduced costs and high availability \cite{nicolas12}. The most popular EEG signals used to control BCI systems are P300 evoked potentials, steady-state visual evoked potentials (SSVEP) and motor imagery (MI) which is the focus of our work. Specifically, MI refers to the imagination of moving certain body parts without actual movement \cite{schuster11}. Different MI tasks result into discriminable patterns observed from the oscillatory activities in the sensorimotor
cortex region of the brain \cite{pfurtscheller99}. Imagination of left hand, right hand, foot and tongue movements are the most investigated MI tasks in the BCI literature \cite{lotte18}.

Handcrafted feature extraction methods coupled with conventional classifiers like Linear Discriminant Analysis (LDA), Support Vector Machines (SVM), Bayesian classifiers, and Nearest Neighbor classifiers have been used in a number of studies for MI task recognition \cite{lotte18}. A widely used approach is to extract and combine band power features from different channel(electrode) signals to capture connectivity patterns among different regions of the sensorimotor cortex and, ultimately, their interaction and engagement with each other. This is thought to play a fundamental role in accomplishing movement imaginations \cite{liu16}. Common spatial patterns (\emph{CSP}) were introduced to this end in \cite{ramoser00} and received a large share of research in the field \cite{blankertz07,lotte10,rivet10,samek13,yger15}, but their effectiveness depended on subject-specific frequency bands. This problem was alleviated by the popular filter bank CSP (\emph{FBCSP}) \cite{ang12} that decomposes the EEG into multiple frequency pass bands prior to spatial filtering, feature selection and classification. This method also won the BCI Competition IV \cite{tangermann12} for 4-class motor imagery recognition (Dataset 2a) and was since used as a reference method for comparison.

Given their effectiveness in other fields \cite{he15,silver16}, deep learning methods, and in particular Convolutional Neural Networks (CNNs)\cite{lecun15}, have the potential to learn both
effective features and classifiers simultaneously from raw EEG
data. Several studies have recently explored deep learning for MI
classification \cite{lu16,schirrmeister17,sturm16,tabar16,lawhern18}. Notably, \cite{schirrmeister17} showed that their \emph{Shallow
ConvNet} (one temporal convolution, one spatial convolution, squaring and mean pooling) could outperform their \emph{Deep ConvNet} (temporal convolution, spatial convolution, then three layers of standard convolution) as well as \emph{FBCSP}. A similar result was achieved by \cite{lawhern18} with \emph{EEGNet}, a compact lightweight network (one temporal convolution, one depthwise convolution, one separable convolution, and a fully connected layer) that compared favorably with \emph{Deep ConvNet} and performed on par with \emph{Shallow ConvNet}. These results indicate that shallow networks having a small number of parameters are beneficial for MI applications that are characterized by very small numbers of training examples because of the difficulty in performing millions or even thousands of mental commands during training sessions.

\begin{figure*}
	\centering
	\includegraphics[width=\textwidth]{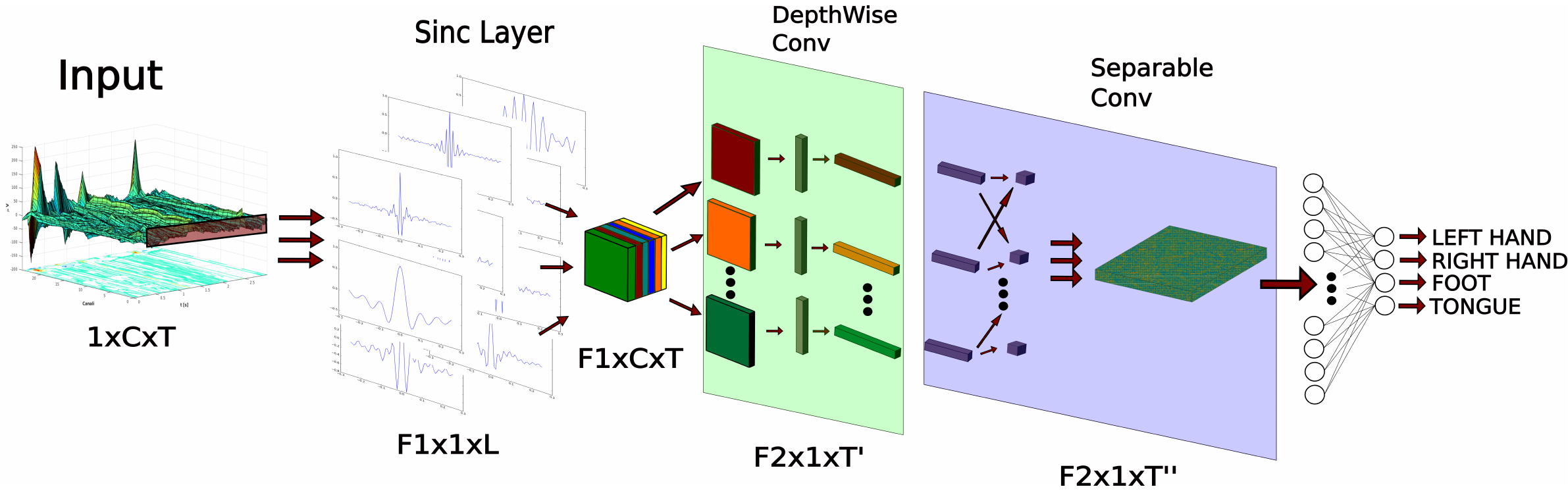}
	\caption{An overview of the proposed \emph{Sinc-EEGNet} architecture.}
	\label{fig:overview}
\end{figure*}

In this paper we propose \emph{Sinc-EEGNet}, a 4-layer CNN architecture that combines the benefits of both EEG frequency band decomposition of classical methods, such as \emph{FBCSP}, and automatic feature learning and extraction of lightweight CNN models, such as \emph{EEGNet}. In particular, the first convolutional layer of our network is restricted to use parameterized sinc functions that implement band pass filters. The subsequent depthwise and separable convolution layers learn a spatial filter and combine the features from the different frequency bands previously selected, which are then inputted to the final classification layer. An overview of the proposed architecture is shown in Fig. \ref{fig:overview}.

\section{Sinc layer}
A standard CNN convolution layer applied on a one-dimensional discrete time-domain signal $s[t]$ performs convolutions with $F$ one-dimensional filters $h_1, ..., h_F$ each having $K$ learnable weights. Conversely, the Sinc layer performs convolutions with $F$ predefined functions $g_1, ..., g_F$ each implementing a learnable bandpass filter $G$ as the difference between two low-pass filters in the frequency domain:
\begin{equation}\label{eq:rectfilter}
G[f] = rect\left(\frac{f}{2f_2}\right)-rect\left(\frac{f}{2f_1}\right)
\end{equation}
where $f_1$ and $f_2>f_1$ are the learnable low and high cutoff frequencies. Using the inverse Fourier transform, the time-domain filter $g$ is obtained as:
\begin{equation}\label{eq:sincfilter}
g[t] = 2f_2sinc(2\pi f_2 t)-2f_1sinc(2\pi f_1 t)
\end{equation}
where the sinc function is defined as $sinc(x)=sin(x)/x$.
The cutoff frequencies are initialized by sampling from a Gaussian distribution with mean and variance equal to $f_s/4$, where $f_s$ represents the
sampling frequency of the input signal. The constraint $f_2>f_1$ is implemented by using in Eq. \ref{eq:sincfilter} the following cutoff frequencies $f_1^{abs}$ and $f_2^{abs}$:
\begin{equation}\label{eq:f1}
f_1^{abs}=|f_1|
\end{equation}
\begin{equation}\label{eq:f2}
f_2^{abs}=f_1 + |f_2-f_1|.
\end{equation}
Because of the discrete approximation of $g$, the resulting bandpass filter is nonideal and may present ripples in the passband and limited
attenuation in the stopband. To alleviate this problem, we multiply $g$ with the popular Hamming window $w$ \cite{dsp05} defined as:
\begin{equation}\label{eq:hamming}
w[t]=0.54-0.46 \cdot \cos\left(\frac{2\pi t}{L}\right)
\end{equation}
where $L$ is the number of discrete samples used to approximate $g$. 
The sinc convolutional layer transforming the input signal $s[t]$ into the band-decomposed output signal $o_1,...,o_F$ is then defined by:
\begin{equation}\label{eq:sinc_conv}
o_i[t]=s[t]*\left(g_i[t]\cdot w[t]\right).
\end{equation}

\section{The Sinc-EEGNet architecture}
The proposed Sinc-EEGNet is a combination and adaptation of the Sinc convolution layer originally proposed by \cite{ravanelli18} for speech recognition with \emph{SincNet}, and \emph{EEGNet} \cite{lawhern18} for what concerns the spatial filtering implemented with depthwise convolution. Specifically, the architecture of Sinc-EEGNet (see Fig. \ref{fig:overview} and Table \ref{tab:architecture}) consists of four blocks described as follows:
\begin{enumerate}
	\item \emph{Sinc Convolution}. The first block takes in input a signal having $C$ channels and $T$ time samples, and performs convolution with $F_1$ sinc filters having $L$ time samples. Compared to the first standard convolution layer used in other CNN architectures such as \emph{EEGNet}, here the sinc filters are explicitly designed to learn the optimal band decomposition for the MI classification task and, when the CNN is trained with data from a single BCI user, this will reflect the peculiarities of the EEG oscillatory activity of that user. Another advantage is the reduced number of parameters, from $K\times F_1$ of the standard convolution to $2\times F_1$ of the sinc convolution. This also implies faster convergence and better generalization capabilities especially when using small training sets as in the case of MI applications. Computational efficiency also is improved since the filters are symmetric, thus the convolution can be performed on one side of the filter and inheriting the result for the other half.  
	\item \emph{Depthwise Convolution}. Similarly to \emph{EEGNet} \cite{lawhern18}, we use a Depthwise Convolution layer \cite{chollet17} of size $(C, 1)$ to learn $D$ spatial filters for each of the $F_1$ inputted feature maps across the channel dimension, for a total of $F_2=D\times F_1$ filters. Combined with the first layer that performs optimal band decomposition, this two-step sequence can be considered a `learnable' version of the well known \emph{FBCSP} \cite{ang12} approach.
	\item \emph{Separable Convolution}. Similarly to \emph{EEGNet}, we summarize each feature map individually using a Depthwise Convolution of size $(1, 16)$, and then merge the outputs using $F_2$ $(1, 1)$ Pointwise Convolutions. This allows optimal combination of the information within and across feature maps. 
	\item \emph{Classification}. The last layer is a fully connected layer that receives the flattened features from the previous layer and maps them to 4 decision classes (left hand, right hand, foot, tongue).

\end{enumerate}

At the end of blocks 1-3 we apply Average Pooling of size $(1, 4)$ for dimensionality reduction, Layer Normalization \cite{ba16}, Dropout regularization \cite{srivastava14}, and CELU activation \cite{barron17}. 
Layer Normalization, as opposed to Batch Normalization \cite{ioffe15} used in other architectures (\emph{EEGNet}, \emph{Deep ConvNet}, \emph{Shallow ConvNet}), calculates the mean and variance across channels instead than batches. This is especially useful for BCI datasets characterized by a high number of channels(electrodes) and small batch sizes resulting from the scarcity of training data. As to the CELU activation, it is an improvement over the ELU activation \cite{clevert15} used in other architectures (\emph{EEGNet}, \emph{Deep ConvNet}, \emph{Shallow ConvNet}) since its derivative does not diverge and it contains both the linear transfer function and ReLU \cite{nair10} activation as special cases.

\ctable[
caption={Sinc-EEGNet architecture, where $C$ = number of channels, $T$ = number of time points, $L$ = number of sinc samples, $F_1$ =
number of temporal filters, $D$ = number of spatial filters, $F_2$ = number of
pointwise filters, and $N$ = number of classes.},
label = tab:architecture,
width = \columnwidth, 
pos = !t,
doinside=\scriptsize]{m{1.0cm}m{3.1cm}m{1.2cm}m{1.0cm}XXm{1.3cm}}
{
}
{ 	\toprule  
	Block & Layer & filters  & size & params & Output  & Activation\\  \toprule 
	\multirow{9}{*}{1}	  &	Input &	 		 &		&		  & $(C,T)$ &	\\ \cmidrule{2-7} 
		  &	Reshape &	 	 &		&		  & $(1, C,T)$ &	\\ \cmidrule{2-7}
		  &	Sinc Convolution & $F_1$& $(1,L)$ & $2\times F_1$ & $(F_1, C,T)$ &	\\ \cmidrule{2-7}
		  &	Average Pooling &       & $(1,4)$ &   			  & $(F_1, C,\frac{T}{4})$ &	\\ \cmidrule{2-7}
		  &	Layer Normalization &   &  		  & $2\times F_1$ & $(F_1, C,\frac{T}{4})$ & CELU	\\ \cmidrule{2-7}
		  &	Dropout &   &  		  &  & $(F_1, C,\frac{T}{4})$ &	\\ \cmidrule{1-7}
	\multirow{6}{*}{2}	 &	Depthwise Convolution & $D\times F_1$& $(C,1)$ & $C\times D\times F_1$ & $(D\times F_1, 1,\frac{T}{4})$ &	\\ \cmidrule{2-7}
		  &	Average Pooling &       & $(1,4)$ &   			  & $(D\times F_1, 1,\frac{T}{16})$ &	\\ \cmidrule{2-7}
		  &	Layer Normalization &   &  		  & $2\times D\times F_1$ & $(D\times F_1, 1,\frac{T}{16})$ & CELU	\\ \cmidrule{2-7}
		  &	Dropout &   &  		  &  & $(D\times F_1, 1,\frac{T}{16})$ &	\\ \cmidrule{1-7}
	 &	Depthwise Convolution & $D\times F_1$& $(1,16)$ & $16\times D\times F_1$ & $(D\times F_1, 1,\frac{T}{16})$ &	\\ \cmidrule{2-7}
	&	Layer Normalization &   &  		  & $2\times D\times F_1$ & $(D\times F_1, 1,\frac{T}{16})$ & CELU	\\ \cmidrule{2-7}
	&	Dropout &   &  		  &  & $(D\times F_1, 1,\frac{T}{16})$ &	\\ \cmidrule{2-7}
	3 &	Pointwise Convolution & $F_2$& $(1,1)$ & $F_2\times(D\times F_1)$ & $(F_2, 1,\frac{T}{16})$ &	\\ \cmidrule{2-7}
	&	Average Pooling &       & $(1,4)$ &   			  & $(F_2, 1,\frac{T}{64})$ &	\\ \cmidrule{2-7}
	&	Layer Normalization &   &  		  & $2\times F_2$ & $(F_2, 1,\frac{T}{64})$ & CELU	\\ \cmidrule{2-7}
	&	Dropout &   &  		  &  & $(F_2, 1,\frac{T}{64})$ &	\\ \cmidrule{1-7}
	\multirow{3}{*}{4}	   	&	Flatten &  &   &  & $F_2\times \frac{T}{64}$ &	\\ \cmidrule{2-7}
	    &	Fully Connected &   &  		  & $N\times F_2\times \frac{T}{64}$  & $N$ & Softmax \\
	\bottomrule \\
}

\section{Experiments}

The EEG data used in this study comes from the BCI Competition IV Dataset 2A \cite{tangermann12}. The data consists of four classes of imagined movements of left and right hands, feet and tongue recorded from 9 subjects during two separate sessions, each composed by 288 trials. The EEG data were originally recorded using $C=22$ Ag/AgCl electrodes(channels), sampled at 250 Hz and bandpass filtered between
0.5 and 100 Hz. We applied a further bandpass filtering to suppress frequencies above 64 Hz and resampled the timeseries to 128 Hz as in \cite{lawhern18}. Z-score standardization was used to normalize the signals within each trial.

EEG data were splitted for training and testing according to three different paradigms:
\begin{enumerate}
	\item \emph{Competition-based}. The training and test sets were the same as indicated in the BCI Competition. This allowed to compare our method with reference methods from the literature that reported their results using the same data split, namely \emph{FBSCP} \cite{ang12}, \emph{Deep ConvNet} \cite{schirrmeister17}, and \emph{Shallow ConvNet} \cite{schirrmeister17} as well as all other participants to the original challenge.
	\item \emph{Within-subject}. For each subject, a dedicated experiment was performed using only data from that subject from the BCI Competition training and test sets. 
	\item \emph{Cross-subject}. For each subject, a dedicated experiment was performed using only data from other subjects from the BCI Competition training set, and only data from that subject from the BCI Competition test set.
\end{enumerate}

\begin{figure*}
	\centering
	\includegraphics[width=\textwidth]{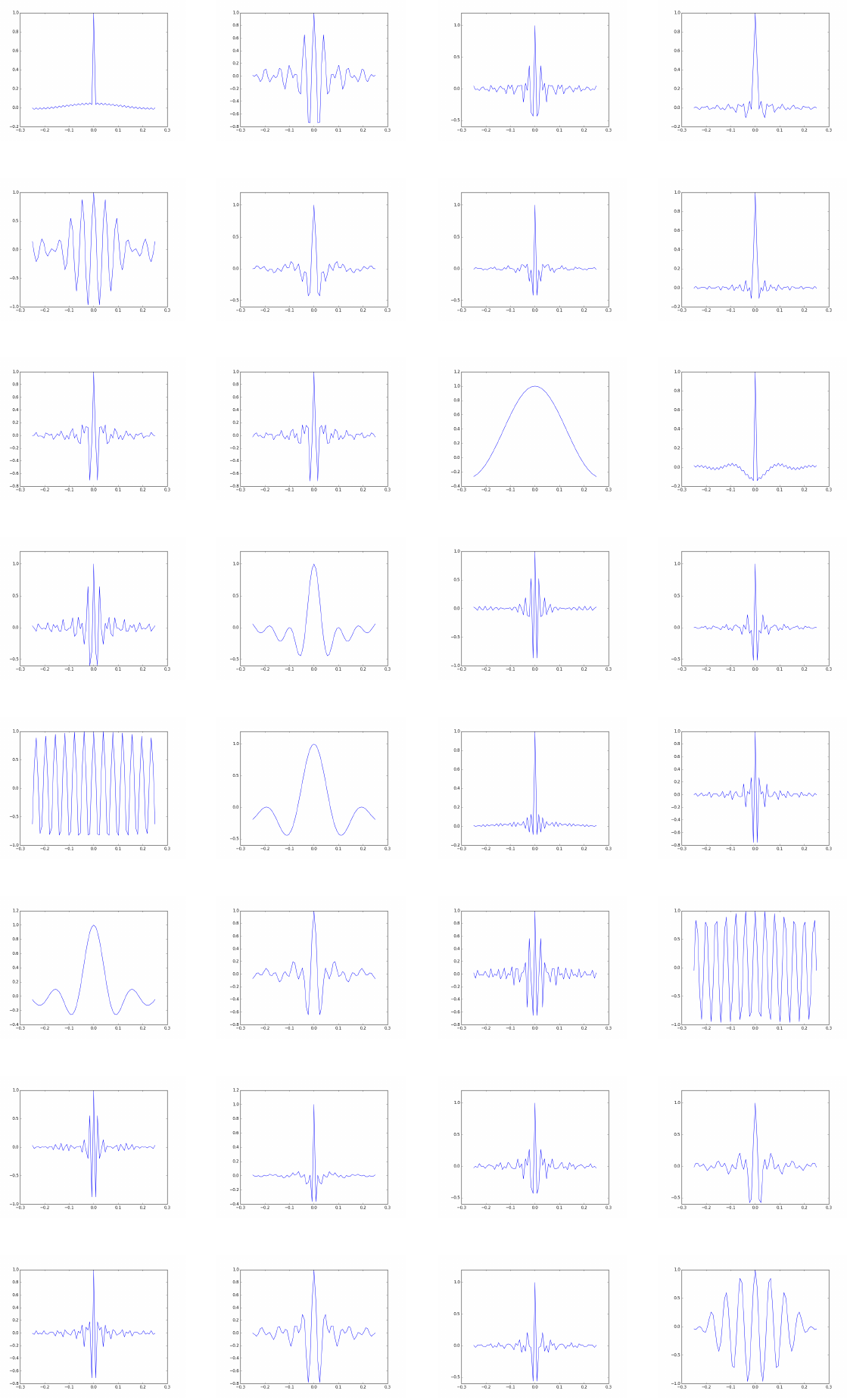}
	\caption{The 32 sinc filters learnt by \emph{Sinc-EEGNet} on the BCI Competition IV Dataset 2A.}
	\label{fig:filters}
\end{figure*}

In all the experiments, we performed a four-class classification using accuracy as the summary measure. In the within- and cross-subject experiments, we also trained and tested an \emph{EEGNet} with $F_1=8$ and $D=2$, which was the best performing CNN reported in \cite{lawhern18}. As to our \emph{Sinc-EEGNet}, we chose $D=2$ for a fair comparison with \emph{EEGNet}, but we set $F_1=32$ since our Sinc layer is specifically designed for frequency band decomposition and thus can benefit from learning a wide variety of bandpass filters. This can be seen in Fig. \ref{fig:filters} that shows 32 distinct filters learnt by \emph{Sinc-EEGNet} in the competition-based experiment. The number of samples $L$ used to discretize the sinc functions was set to $64$ that resulted from a trade-off between approximation precision and computational complexity.

All the CNNs were trained using backpropagation and Adam optimizer \cite{kingma14} with weight updates that proceeded in batches of $20$ samples for $100$ epochs. The base learning rate was set to $10^{-3}$. Momentum and weight decay were set respectively to $0.9$ and $2\times10^{-2}$. Following \cite{lawhern18}, for the Dropout layers we chose $p=0.5$
for within-subject experiments, and $p=0.25$ for competition-based and cross-subject experiments that used more training data and thus required less regularization. The loss function was categorical cross-entropy. 

\ctable[
caption={Comparison of classification accuracies between our method and reference methods on the BCI Competition IV-2A.},
label = tab:results,
pos = !t,
doinside=\scriptsize]{m{4.0cm}m{2.0cm}}
{
}
{ 	\toprule  
	Method & Accuracy\\  \toprule 
	\emph{FBCSP} & $68.0\%$ \\ \cmidrule{1-2}
	\emph{Deep ConvNet} & $70.9\%$ \\ \cmidrule{1-2}
	\emph{Shallow ConvNet} & $73.7\%$ \\ \cmidrule{1-2}
	\emph{Sinc-EEGNet} & $75.39\%$ \\ 
	\bottomrule \\
}

\section{Results}
The comparison between \emph{Sinc-EEGNet} and the reference methods from the literature on the competition-based data split are reported in Table \ref{tab:results}. Remarkably, \emph{Sinc-EEGNet} outperforms all other methods in terms of accuracy and sets a new state-of-the-art on the BCI Competition IV-2A with an accuracy of $75.39\%$ that improves \emph{FBCSP} by $17.39\%$. As to the within- and cross-subject experiments, \emph{EEGNet} yielded an average accuracy of $60.99\%$ and $58.75\%$, respectively, and \emph{Sinc-EEGNet} of $70.56\%$ and $58.98\%$, respectively. Also in this case, our method exhibited superior performance, with an improvement of almost $10\%$ accuracy in the more practically adopted within-subject classification.

\section{Conclusions}
In this work we proposed \emph{Sinc-EEGNet}, a lightweight convolutional neural network for EEG-BCI-based motor imagery classification that learns optimal band decomposition and spatial filtering, mimicking the behavior of the well-known \emph{FBCSP} but learning the filters directly from the raw EEG data. Our method outperformed reference methods from the literature, including \emph{FBCSP} and \emph{EEGNet}, on the publicly available BCI Competition IV-2A dataset. To the best of our knowledge, this is the first work that validated the use of learnable bandpass filters in the first layer of a CNN for EEG signal classification. Future work will investigate alternative frequency filters, such as Difference of Gaussian (DoG) filter, that are less subject to discrete approximation issues, and architecture variants that explore different spatial filtering and feature map combination approaches.

%
%
%
\bibliographystyle{splncs04}
\bibliography{refs}
\end{document}